# Thermodynamic perturbation theory for associating fluids with small bond angles: effects of steric hindrance, ring formation and double bonding

Bennett D. Marshall and Walter G. Chapman.[1]
Department of Chemical and Biomolecular Engineering
Rice University
6100 S. Main
Houston, Texas  77005

## Abstract

We develop the first comprehensive approach to model associating fluids with small bond angles using Wertheim's perturbation theory. We show theoretically and through monte carlo simulations that as bond angle is varied various modes of association become dominant. The new theory is shown to be in excellent agreement with monte carlo simulation for the prediction of the internal energy, pressure and fractions in rings, chains, double bonded over the full range of bond angles.

---

[1] Author to whom correspondence should be addressed.  Tel: (1) 713.348.4900.  Fax: (1) 713.348.5478.  Email:  wgchap@rice.edu.

**I: Introduction**

The statistical mechanics of associating fluids was the focus of much research in the 1980's -1990's as researchers sought to develop equations of state for hydrogen bonding fluids based on primitive models of the hydrogen bond. [1-8] The difficulty in modeling hydrogen bonding fluids arises from the anisotropic nature of the hydrogen bond as well as the fact that hydrogen bonds saturate. In recent years these primitive models have gained recognition among researchers studying patchy colloids.[9] Patchy colloids have discrete attractive patches resulting in orientation dependant potentials which can be manipulated by varying the number, size, strength and location of these patches giving researchers the ability to program self assembly into pre-determined structures.[10, 11] This control may allow for the design of a new generation of functional materials.[12] In addition, it has been shown that the properties of patchy colloids can be varied to yield exotic phase behavior such as empty liquids[13] and re-entrant phase behavior of network fluids.[14]

To develop a theory capable of modeling associating fluids (or patchy colloids) one must be able to account for the fact that bonding at one patch can block other colloids from approaching this patch to form another attraction bond. That is, for small patch size, the patch will saturate in the sense that only a single attractive bond per patch is allowed. This was the problem tackled by Wertheim[4-8], in the context of a primitive model for hydrogen bonding, who converted statistical mechanics into a multi - density formalism where each bonding state was treated as an independent species. In its general form Wertheim's theory provides an exact solution in terms of a multi – density cluster expansion.[6] The theory is intractable in its most general form; however, vast simplification occurs if we restrict attractions such that only one bond per patch is allowed. For small to moderate patch size this restriction is reproduced naturally and exactly in Wertheim's theory. Recently progress has been made on relaxing this restriction and allowing for larger patch sizes.[15-18]

Conceptually, Wertheim's theory can be thought of as a virial expansion in association (attractions for patchy colloids). Restricting our attention to one patch colloids with a single bond per patch (such that there are only spheres and associated dimers), the contributions to the free energy can be described as follows: 1. the zeroth order contribution accounts for all hard sphere interactions, 2. the first order contribution accounts for the association of two spheres into a dimer and the interaction of this dimer with the hard sphere reference fluid, 3. the second order contribution accounts for the association of four spheres into two dimers and the interactions of these two dimers with each other and the hard sphere reference fluid etc… The zeroth order contribution is known from the reference system equation of state and all graphs in first order contribution can be condensed into a single graph containing the pair correlation function of the hard sphere reference fluid in the form of a perturbation theory. In all applications of Wertheim's theory all contributions containing more than one path of attraction bonds (dimers in the one patch case) are neglected. This is the single chain approximation.

Restricting association to one bond per patch and enforcing the single bonding condition results in significant simplification of the theory. For the one patch case the path of approximation ends here; however, in the two patch case other levels of approximation exist. In the two patch case in the single chain approximation the free energy in Wertheim's perturbation theory will contain contributions for the association of a single pair of colloids (first order perturbation theory TPT1), a triplet of colloids (TPT2) etc… In TPT1 only the pair correlation function of the hard sphere reference fluid is needed and chains of any size can be modeled to a reasonable degree of accuracy[3, 8]. Since TPT1 only accounts for association between pairs of colloids, the angle between patches or bond angle $\alpha_{AB}$ does not enter the theory. This is accurate for large $\alpha_{AB}$ but for small $\alpha_{AB}$ it is possible that bonding at one patch can block bonding at the other. To account for blocking effects Wertheim carries out a resumed perturbation theory[8] which in first order requires only the pair correlation function of the hard sphere reference system and a blockage parameter which accounts for the decrease in bonding volume of one patch caused by association of the other.

Also neglected in TPT1, is the possibility of double bonding of colloids (a pair of colloids share two attraction bonds) as well as the possibility of ring formation. The formation of double bonded dimers was a problem initially tackled by Sear and Jackson[19] (SJ) who included the additional contribution for the double bonded dimer. The theory includes a geometric quantity which accounts for the probability that two colloids will be oriented such that double bonding can occur. This quantity was never explicitly evaluated and was treated as a parameter allowing only qualitative comparisons to be made. The ability of colloids to double bond will be strongly dependant on the angle between patches $\alpha_{AB}$.

It was also SJ who were the first to introduce contributions for the association of colloids with two patches into rings.[20] In this approach the associated rings were treated ideally such that non-adjacent neighbors along the ring can overlap. The probability that a chain of colloids was in a valid ring state was approximated by the expression of Treolar[21] for the distribution of the end to end vector in a polymer chain. In this approach any dependence on $\alpha_{AB}$ is neglected when in reality $\alpha_{AB}$ plays a dominant role in determining if association into rings will occur. A recent study using lattice simulations has shown that ring formation is strongly dependent on $\alpha_{AB}$.[22] For instance, it is impossible to form 4-mer rings (and satisfy the one bond per patch condition) from 2 patch colloids with patches at opposite ends of the colloid $\alpha_{AB} = 180°$; however, decreasing $\alpha_{AB}$ to 90° this type of ring would be possible. Tavares et al.[23] explored the possibility of ring formation in 2 patch colloid fluids with $\alpha_{AB} = 180°$ by extending the approach of SJ[20] and found that to achieve appreciable ring formation the parameters of the interaction potential had to be chosen such that the one bond per patch condition would be violated. To correct for this in the simulations they used a model which restricts bonding to at most one bond per patch.[24]

In this work we wish to extend Wertheim's perturbation theory to model 2 patch fluids with small (or large) bond angles $\alpha_{AB}$. Our goal is to derive a single theory which will be accurate over the full range of $\alpha_{AB}$. To accomplish this we will combine and extend the resummed perturbation theory of Wertheim[8], theory for double bonded dimers of SJ[19] and a modified version of the approach of SJ[20] for ring formation. We explicitly include the dependence on bond angle $\alpha_{AB}$ in each contribution of the theory and evaluate all required geometric integrals rigorously. To test the new theory we perform new monte carlo simulations to determine the effect of $\alpha_{AB}$ on the fractions of colloids associated into chains, rings, double bonded dimers as well as the effect of $\alpha_{AB}$ on internal energy and pressure. The theory is found to be in excellent agreement with simulation.

For the patchy colloid community the results of this paper can be used a tool to aid in the design of colloids which self assemble into predetermined structures. The case of double bonding of colloids may be most relevant to the situation of DNA tethered colloids. Far from being restricted to anisotropic colloids, the results presented in this paper can also be applied to describe ring formation, double bonding and steric hindrance in hydrogen bonding fluids.

The paper is organized as follows. In section II the general theory will be derived in the framework of Wertheim's perturbation theory. Section III gives a brief description of the simulation method used and section IV gives an extensive comparison between monte carlo simulation and theory predictions. Finally in section V we give conclusions and discussion of future work.

**II: Theory**

In this section the theory will be developed for colloids of diameter $\sigma$ with an *A* patch and a *B* patch with the centers of the patches having a bond angle $\alpha_{AB}$ in relation to each other. The size of the patches is controlled by the angle $\beta_c$ which defines the solid angle of the patch $2\pi(1 - \cos\beta_c)$. A diagram of this type of colloid can be found in Fig. 1. The potential of interaction between two colloids is given by the sum of a hard sphere potential $\phi_{HS}(r_{12})$ and orientation dependant attractive patchy potential $\phi_{AB}(12)$

$$\phi(12) = \phi_{HS}(r_{12}) + \phi_{AB}(12) + \phi_{BA}(12) \tag{1}$$

The notation $(1) \equiv (\vec{r}_1, \Omega_1)$ represents the position $\vec{r}_1$ and orientation $\Omega_1$ of colloid 1 and $r_{12}$ is the distance between the colloids. Here we follow Bol[25] and Chapman et al.[3] who employed a potential for conical association sites

$$\phi_{AB}(12) = \begin{cases} -\varepsilon_{AB}, & r_{12} \leq r_c \text{ and } \beta_{A1} \leq \beta_c \text{ and } \beta_{B2} \leq \beta_c \\ 0 & \text{otherwise} \end{cases} \tag{2}$$

which states that if colloids 1 and 2 are within a distance $r_c$ of each other and each colloid is oriented such that the angles between the site orientation vectors and the vector connecting the two segments, $\beta_{A1}$ for colloid 1 and $\beta_{B2}$ for colloid 2, are both less than the critical angle $\beta_c$, the two sites are considered bonded and the energy of the system is decreased by a factor $\varepsilon_{AB}$. In this work there are no attractive interactions between like patches, that is $\varepsilon_{AA} = \varepsilon_{BB} = 0$. To ensure that each patch can only bond once we choose $r_c = 1.1\sigma$ and $\beta_c = 27°$. Kern and Frenkel[26] where the first to realize the potential given by Eqns. (1) – (2) provided an excellent description of the interactions between patchy colloids.

We will develop the equation of state using Wertheim's theory.[4-8] For the 2 patch case the Helmholtz free energy is given by

$$\frac{A - A^{HS}}{Vk_BT} = \rho \ln\left(\frac{\rho_o}{\rho}\right) - \sigma_A - \sigma_B + \frac{\sigma_A \sigma_B}{\rho_o} + \rho - \frac{\Delta c^{(o)}}{V} \qquad (3)$$

Where $A^{HS}$ is the free energy of the hard sphere reference system, $\rho$ is the total number density, $\rho_o$ is the density of monomers, $\sigma_A = \rho_A + \rho_o$ where $\rho_A$ is the density of colloids which are bonded at patch $A$ (a similar relation exist for $\sigma_B$) and $V$ is volume. The term $\Delta c^{(o)}$ is the associative contribution to the fundamental graph sum which encodes all of the attractive interactions. For colloids with small bond angles we will have to account for chain formation $\Delta c_{ch}^{(o)}$, association into double bonded dimers $\Delta c_d^{(o)}$ and lastly rings of associated colloids $\Delta c_{ring}^{(o)}$ giving the graph sum

$$\Delta c^{(o)} = \Delta c_{ch}^{(o)} + \Delta c_d^{(o)} + \Delta c_{ring}^{(o)} \qquad (4)$$

The various modes of association are outlined in Fig.2. In the work on double bonded dimers by SJ[19] the possibility of ring formation was not included in the formulation. As will be shown ring formation becomes very important at small bond angles. For the chain contribution we use the first order resummed perturbation theory RTPT1 of Wertheim[8] to account for the fact that at small bond angles, association at one patch can block association at the other

$$\frac{\Delta c_{ch}^{(o)}}{V} = \frac{\sigma_A \sigma_B \kappa f_{AB} \xi}{1 + (1 - \Psi)\kappa f_{AB} \rho_o \xi} \qquad (5)$$

where $f_{AB} = \exp(\varepsilon_{AB}/k_BT) - 1$ is the association Mayer function, $\kappa = (1 - \cos\beta_c)^2/4$ is the probability that two monomers are oriented such that a certain patch on one colloid can bond to a certain patch on the other, and $\xi$ is given by

$$\xi = 4\pi \int_\sigma^{r_c} r^2 g_{HS}(r) dr \qquad (6)$$

where $g_{HS}(r)$ is the hard sphere reference system pair correlation function. Since the range of the integration in Eq. (6) is small, $r_c = 1.1\sigma$, it is common practice to use a Taylor's series expansion of $g_{HS}(r)$ around the value at hard sphere contact $g_{HS}(\sigma)$ such that

$$g_{HS}(r) = g_{HS}(\sigma) + \left.\frac{\partial g_{HS}(r)}{\partial r}\right|_{r=\sigma}(r-\sigma) \qquad (7)$$

However, in the current work this approximation of $g_{HS}(r)$ will prove very inconvenient in the evaluation of $\Delta c_{ring}^{(o)}$. As an alternative we employ the fact that in the bonding range $\{\sigma \leq r \leq r_c\}$ the following relation holds true to an excellent approximation[2]

$$r^p g_{HS}(r) = \sigma^p g_{HS}(\sigma) \qquad (8)$$

Where $p$ is a density dependent quantity which we obtain by fitting Eq. (8) to the analytical solution for $g_{HS}(r)$ of Chang and Sandler[27]. The results can be represented by the simple polynomial $p = 17.87\eta^2 + 2.47\eta$, where $\eta = \pi\rho\sigma^3/6$ is the packing fraction. Combining (6) and (8) we obtain the simple result

$$\xi = 4\pi\sigma^3 g_{HS}(\sigma)\left(\frac{(r_c/\sigma)^{3-p} - 1}{3-p}\right) \qquad (9)$$

The last term to consider in $\Delta c_{ch}^{(o)}$ is the blockage integral $\Psi$ which accounts for the fact that as the bond angle is decreased bonding at one patch will decrease the available bond volume of the other patch due to steric hindrance. Wertheim developed RTPT1 in the context of spheres which bond at contact with 2 glue spots *A* and *B*. In Wertheim's treatment once spheres formed a bond they were stuck and would not rattle around in the bond volume defined by Eqn. (1). In the current work $\alpha_{AB}$ defines the angle between patch centers which we call the bond angle; for a given $\alpha_{AB}$ the actual angle $\alpha'_{AB}$ between the first and third spheres in an associated linear triatomic cluster can vary in the range $\alpha_{AB} - 2\beta_c \leq \alpha'_{AB} \leq \alpha_{AB} + 2\beta_c$. In Wertheim's analysis of RTPT1 this was not the case; there was no rattling in the bond volume meaning $\alpha'_{AB} = \alpha_{AB}$. To account for bond flexibility Wertheim introduced normalized bond angle distribution functions $\zeta(\alpha_{AB})$. For hard spheres with glue spot bonding Wertheim found

$\Psi = 1 - L = \int_{\pi/3}^{\pi} \zeta(\alpha_{AB}) \sin(\alpha_{AB}) d\alpha_{AB}$; which is simply the fraction of $\alpha_{AB}$ states which will not result in hard sphere overlap when both glue spots are bonded. Our case here is somewhat different since we do not have glue spot bonding. We set $\alpha_{AB}$, $r_c$ and $\beta_c$, not $\zeta(\alpha_{AB})$; however the interpretation of $\Psi$ is similar. In our case $\Psi$ is the ratio of the number of states where three colloids associate to form a linear triatomic cluster in which there is no hard sphere overlap between the unbounded pair to the number of states if there were no steric interference and the patches were independent. For the model considered here this fraction $\Psi$ is approximated by the integral

$$\Psi = \frac{\int_0^{2\pi}\int_0^{\beta_c}\int_\sigma^{r_c}\int_0^{2\pi}\int_0^{\beta_c}\int_\sigma^{r_c} dr_{12} r_{12}^2 d\theta_{12} \sin\theta_{12} d\phi_{12} dr_{13} r_{13}^2 d\theta_{13} \sin\theta_{13} d\phi_{13} e_{HS}(r_{23})}{\left(\int_0^{2\pi}\int_0^{\beta_c}\int_\sigma^{r_c} dr_{12} r_{12}^2 d\theta_{12} \sin\theta_{12} d\phi_{12}\right)\left(\int_0^{2\pi}\int_0^{\beta_c}\int_\sigma^{r_c} dr_{13} r_{13}^2 d\theta_{13} \sin\theta_{13} d\phi_{13}\right)} \qquad (10)$$

Where in Eq. (10) colloid 2 and 3 are both bonded to colloid 1, $\theta_{12}$ is the polar angle that the vector $\vec{r}_{12} = \vec{r}_2 - \vec{r}_1$ makes in a coordinate system centered on colloid 1 whose $z$ axis lies on the site vector $\vec{r}_A$ of colloid 1 and $\phi_{12}$ is the corresponding azimuthal angle. The angles $\theta_{13}$ and $\phi_{13}$ are similarly defined with respect to the site vector $\vec{r}_B$ of colloid 1. The $e_{HS}$ prevents hard sphere overlap between the colloids in the associated cluster and is given by $e_{HS}(r) = H(r-\sigma)$ where $H$ is the Heaviside step function. For the case of total blockage of one patch by the other the reference system $e_{HS}(r_{23}) = 0$ for all configurations of the cluster resulting in $\Psi = 0$. When the bond angle is sufficiently large that the two patches are independent the reference system $e_{HS}(r_{23}) = 1$ for all configurations resulting in $\Psi = 1$.

When the condition $\alpha_{AB} - 2\beta_c < 0$ holds true it is possible for double bonding of colloids to occur; that is, according to the potential given by Eq. (2), when the vector connecting the centers of the two colloids $\vec{r}_{12}$ passes through both patches on both colloids and $r_{12} \leq r_c$, the two colloids are considered to be double bonded. This situation is depicted in Fig. 3. The contribution to the graph sum which accounts for double bonded dimers is given by SJ (we introduce different constants but the result is the same)[19]

$$\Delta c_d^{(o)}/V = \rho_o^2 f_{AB}^2 \xi I_d / 2 \qquad (11)$$

Where $I_d$ is the probability that two colloids are oriented such that double bonding can occur. In the work of SJ[19] $I_d$ was not defined in this way and was not explicitly evaluated; instead it was written in terms of a parameter which allowed only qualitative discussion and could not explicitly be compared to simulation. With the identification of $I_d$ as the probability two colloids are oriented for double bonding and the definition of the patchy potential in Eq. (2) we can easily evaluate $I_d$ as follows. Consider the two colloids in Fig. 3 whose bond angle satisfies the condition $\alpha_{AB} - 2\beta_c \leq 0$. For one colloid to be in an orientation to double bond to the other, the vector connecting the centers of the two colloids must pass through the area of surface where the two patches overlap. This is the dashed outline area in Fig. 3. If all orientations of a colloid are equally likely, the probability of this occurring will simply be the ratio of the surface area of patch overlap to the total surface area of the sphere $S_{AB}/4\pi$ where $S_{AB}$ is the solid angle of the overlap of the two patches. This solid angle $S_{AB}$ is simply the solid angle of the intersection of 2 cones of apex angle $\beta_c$ which share a common origin whose axes are at an angle of $\alpha_{AB}$ to each other. For association to occur both colloids must be oriented correctly, so we square the single colloid probability to obtain

$$I_d = \begin{cases} 0 & for \ \alpha_{AB} - 2\beta_c \geq 0 \\ S_{AB}^2 / 16\pi^2 & otherwise \end{cases} \qquad (12)$$

The solid angle $S_{AB}$ has been solved elsewhere[28] and is given by

$$S_{AB} = 4\cos^{-1}\left(\frac{\sin\gamma_{AB}}{\sin\beta_c}\right) - 4\cos(\beta_c)\cos^{-1}\left(\frac{\tan\gamma_{AB}}{\tan\beta_c}\right) \tag{13}$$

where $\gamma_{AB}$ is obtained through the relation $\gamma_{AB} = \tan^{-1}((1-\cos\alpha_{AB})/\sin\alpha_{AB})$.

Finally, we must account for the possibility of rings of associated colloids with the contribution

$$\Delta c_{ring}^{(o)} = \sum_{n=3}^{\infty} \Delta c_n^{ring} \tag{14}$$

where $\Delta c_n^{ring}$ is the contribution for rings of size $n$. The contributions $\Delta c_n^{ring}$ are given by

$$\Delta c_n^{ring}/V = \frac{\rho_o^n}{n(8\pi^2)^{n-1}} \int f_{AB}(1 2)\ldots f_{AB}(n-1,n) f_{AB}(1,n) g_{HS}(\vec{r}_1\ldots\vec{r}_n) d\vec{r}_2 d\Omega_2 \ldots d\vec{r}_n d\Omega_n \tag{15}$$

where $f_{AB}(1,2) = \exp(-\phi_{AB}(12)/k_B T) - 1$. Equation (15) is more general than the ring graph of Sear and Jackson[20] with the introduction of the $n$ - body correlation function of the hard sphere reference system $g_{HS}(\vec{r}_1\ldots\vec{r}_n)$ which we approximate as

$$g_{HS}(\vec{r}_1\ldots\vec{r}_n) = \prod_{\substack{\text{bonded pairs}\\\{i,j\}}} y_{HS}(r_{ij}) \prod_{\substack{\text{all pairs}\\\{l,k\}}} e_{HS}(r_{lk}) \tag{16}$$

The superposition given by Eq. (16) gives a $g_{HS}(r_{ij}) = y_{HS}(r_{ij}) e_{HS}(r_{ij})$ to each pair of colloids which share an association bond and an $e_{HS}(r_{lk})$ to each unbounded pair which serves to prevent hard sphere overlap between non-adjacent spheres in the ring.

In the approach taken by SJ[21], Eq. (16) is replaced by a simple linear superposition of pair correlation functions and the integral in Eq. (15) is approximated as $\Delta c_n^{ring}/V = (f_{AB}\rho_o g_{HS}(\sigma)K)^n W_{n-1}/n$ where $K = 4\pi\sigma^2(r_c - \sigma)\kappa$ and $W_{n-1}$ is the probability that, in a freely jointed chain, the first and last sphere in the chain are in contact and is obtained using the expression of Treolar[21]. The SJ approximation of the ring integral is not useful in our current approach because the effect of bond angle has not been included. For instance, in the SJ approximation there will be a significant probability colloids which have a bond angle of 180° will associate into triatomic rings while in reality this is geometrically impossible.

In this work we treat Eq. (15) in a more general way which allows for the inclusion of bond angle dependence. First, we note that with the potential given by Eq. (2) we can rewrite the association Mayer functions as

$$f_{AB}(12) = f_{AB}\frac{\phi_{AB}(12)}{-\varepsilon_{AB}} \tag{17}$$

Using the approximation Eq. (8) with the fact that for $r \geq \sigma$, $g_{HS}(r) = y_{HS}(r)$ allows us to rewrite $\Delta c_n^{ring}$ as

$$\Delta c_n^{ring}/V = \frac{(f_{AB}\rho_o g_{HS}(\sigma)K)^n}{n\sigma^3}I_r^{(n)} \tag{18}$$

Where the ring integral $I_r^{(n)}$ is given by

$$I_r^{(n)} = \frac{\sigma^3}{(8\pi^2)^{n-1}K^n}\int \prod_{\substack{bonded\ pairs\\\{i,j\}}}\left(\frac{\sigma^p\phi_{AB}(i,j)}{r_{ij}^p\ -\varepsilon_{AB}}\right)\prod_{\substack{all\ pairs\\\{l,k\}}}e_{HS}(r_{lk})d\vec{r}_2d\Omega_2\ldots d\vec{r}_nd\Omega_n \tag{19}$$

The probability distribution function of a ring of size *n* is in a configuration (1...*n*) is given by

$$P_r^{(n)}(1...n) = \frac{\prod_{\substack{bonded\ pairs \\ \{i,j\}}} \left(\frac{\phi_{AB}(i,j)}{-\varepsilon_{AB}}\right) \prod_{\substack{all\ pairs \\ \{l,k\}}} e_{HS}(r_{lk})}{\Theta^{(n)}} \quad (20)$$

Where $\Theta^{(n)}$ is the normalization constant given by

$$\Theta^{(n)} = \int \prod_{\substack{bonded\ pairs \\ \{i,j\}}} \left(\frac{\phi_{AB}(i,j)}{-\varepsilon_{AB}}\right) \prod_{\substack{all\ pairs \\ \{l,k\}}} e_{HS}(r_{lk}) d\vec{r}_2 d\Omega_2 ... d\vec{r}_n d\Omega_n \quad (21)$$

Combining (19) – (21) we obtain

$$I_r^{(n)} = \Gamma^{(n)} \left\langle \prod_{\substack{bonded\ pairs \\ \{i,j\}}} \left(\frac{\sigma^p}{r_{ij}^p}\right) \right\rangle \quad (22)$$

Where

$$\Gamma^{(n)} = \frac{\sigma^3 \Theta^{(n)}}{(8\pi^2)^{n-1} K^n} \quad (23)$$

and $\langle \ \rangle$ represents an average over the distribution function Eq. (20). Since $P_r^{(n)}$ is only nonzero when there is no hard sphere overlap we can accurately approximate this average as

$$\left\langle \prod_{\substack{bonded\ pairs \\ \{i,j\}}} \left(\frac{\sigma^p}{r_{ij}^p}\right) \right\rangle \approx \frac{2^{np}}{(r_c/\sigma + 1)^{np}} \quad (24)$$

which states that on average each colloid pair should be approximately located in the middle of the range $\{\sigma \leq r \leq r_c\}$. Using this approach $\Delta c_n^{ring}$ is explicitly dependent on bond angle through the normalization constant $\Theta^{(n)}$ which must be evaluated numerically.

For the two patch case three densities describe all possible bonding states of the colloid: the monomer density $\rho_o$, density of colloids bonded at patch A (or equivalently B) $\rho_A = \rho_B$ and the density of colloids bonded at patches A and B $\rho_{AB}$. In the graphical formalism of Wertheim $\rho_A$ and $\rho_{AB}$ are related to $\Delta c^{(o)}$ by the following relations[6-8]

$$\frac{\rho_A}{\rho_o} = c_A = \frac{\partial \Delta c^{(o)}/V}{\partial \sigma_B} \qquad (25)$$

and

$$\frac{\rho_{AB}}{\rho_o} = c_{AB} + c_A c_B \qquad (26)$$

Where

$$c_{AB} = \frac{\partial \Delta c^{(o)}/V}{\partial \rho_o} \qquad (27)$$

and $c_A = c_B$. The free energy in Eq. (3) is constructed such that free energy minimization corresponds to these relations.[6] Using Eqns. (4) – (5) and (25) we obtain the density of colloids bonded at patch A as

$$\frac{\rho_A}{\rho_o} = \frac{\sigma_A}{\rho_o} - 1 = \frac{\sigma_A \kappa f_{AB} \xi}{1 + (1-\Psi)\kappa f_{AB}\rho_o \xi} \qquad (28)$$

The density of colloids bonded twice is given by Eqns. (4) – (5), (11), (18) and (26)

$$\frac{\rho_{AB}}{\rho_o} = \frac{\partial}{\partial \rho_o} \frac{\sigma_B \sigma_A \kappa f_{AB} \xi}{1+(1-\Psi)\kappa f_{AB}\rho_o \xi} + \left(\frac{\sigma_A \kappa f_{AB} \xi}{1+(1-\Psi)\kappa f_{AB}\rho_o \xi}\right)^2 + \rho_o f_{AB}^2 \xi I_d + \sum_{n=3}^{\infty} \frac{(f_{AB}\rho_o g_{HS}(\sigma)K)^n}{\rho_o \sigma^3} I_r^{(n)}$$

(29)

The density of colloids bonded twice must satisfy the relation

$$\rho_{AB} = \rho_{2c} + \rho_d + \sum_{n=3}^{\infty} \rho_n^{ring}$$

(30)

Where $\rho_{2c}$ is the density of colloids which are bonded at both patches in a linear chain, $\rho_d$ is the density of colloids in double bonded dimers and $\rho_n^{ring}$ is the density of colloids in rings of size $n$. The first two terms on the right hand side (RHS) of Eq. (29) correspond to $\rho_{2c}$ and can be simplified as

$$\frac{\rho_{2c}}{\rho_o} = \Psi \left(\frac{\sigma_A \kappa f_{AB} \xi}{1+(1-\Psi)\kappa f_{AB}\rho_o \xi}\right)^2$$

(31)

For complete blockage of one patch by the other $\Psi \to 0$, resulting in $\rho_{2c} \to 0$, while for independent patches $\Psi \to 1$ and the TPT1 result is obtained. Likewise, the third term on the RHS of Eq. (29) corresponds to $\rho_d$

$$\frac{\rho_d}{\rho_o} = \rho_o f_{AB}^2 \xi I_d$$

(32)

and the fourth term on the RHS of Eq. (29) gives the sum of the densities of colloids in rings of size $n$

$$\frac{\rho_n^{ring}}{\rho_o} = \frac{\left(f_{AB}\rho_o g_{HS}(\sigma)K\right)^n}{\rho_o \sigma^3} I_r^{(n)} \qquad (33)$$

The total density is given as the sum over all of the bonding states of the colloids

$$\rho = \rho_o + 2\rho_A + \rho_{AB} \qquad (34)$$

Since $\rho$ is known Eqns. (28) and (34), with $\rho_{AB}$ given by Eq. (30), provide a closed set of equations to solve for $\rho_o$ and $\rho_A$ from which $\rho_{2c}$, $\rho_d$ and $\rho_n^{ring}$ immediately follow.

What we have done is solve for the density of each bonding state of the colloid in a self consistent manner. To simplify the free energy given in Eq. (3) we relate the various contributions to $\Delta c^{(o)}$ to their respective density. Comparing Eqns. (5) and (28) we see

$$\frac{\Delta c_{ch}^{(o)}}{V} = \frac{\sigma_A \sigma_B}{\rho_o} - \sigma_A \qquad (35)$$

Comparing Eqns. (11) and (32)

$$\Delta c_d^{(o)}/V = \rho_d/2 \qquad (36)$$

And finally comparing Eqns. (18) and (33)

$$\Delta c_n^{ring}/V = \rho_n^{ring}/n \qquad (37)$$

Combining Eqns. (3) and (35) – (37) we obtain the final form of the Helmholtz free energy

$$\frac{A - A^{HS}}{Nk_B T} = \ln X_o + 1 - X_A - \frac{X_d}{2} - \sum_{n=3}^{\infty} \frac{X_n^{ring}}{n} \tag{38}$$

Where we have introduced the fractions $X_A = \sigma_B / \rho$ is the fraction of colloids not bonded at patch A, $X_d = \rho_d / \rho = f_{AB}^2 \xi \rho I_d X_o^2$ is the fraction of colloids in double bonded dimers and $X_n^{ring} = \rho_n^{ring} / \rho = (f_{AB} g_{HS}(\sigma) K)^n \rho^{n-1} I_r^{(n)} X_o^n / \sigma^3$ is the fraction of colloids associated in rings of size $n$.

The chemical potential is evaluated from Eq. (38) as

$$\frac{\mu - \mu^{HS}}{k_B T} = \ln X_o - \frac{1}{2}\left(X_1 + \frac{1}{2}\frac{X_1^2}{X_o}\Psi + X_d\right)\eta \frac{\partial \ln \xi}{\partial \eta} - \left(\eta \frac{\partial \ln g_{HS}(\sigma)}{\partial \eta} - \ln\left(\frac{r_c/\sigma + 1}{2}\right)\eta \frac{\partial p}{\partial \eta}\right)\sum_{n=3}^{\infty} X_n^{ring} \tag{39}$$

where $X_1 = 2(X_A - X_o)$ is the fraction of spheres bonded once at either patch. Equations (38) and (39) require the fractions $X_A$ and $X_o$ which are obtained by solving the following set of equations. From Eq. (28)

$$X_A = \frac{X_o}{1 - \lambda X_o} \tag{40}$$

Where $\lambda$ depends on the monomer fraction

$$\lambda \equiv \frac{\rho \kappa f_{AB} \xi}{1 + (1 - \Psi)\kappa f_{AB} \rho X_o \xi} \tag{41}$$

From Eqns. (25) – (27) and (34)

$$\frac{1}{X_o} = (1 + X_A \lambda)^2 + \rho X_o f_{AB}^2 \xi I_d + \sum_{n=3}^{\infty} \frac{(f_{AB} g_{HS}(\sigma) K)^n}{\sigma^3}(\rho X_o)^{n-1} I_r^{(n)} - (1 - \Psi)(X_A \lambda)^2 \tag{42}$$

Combining Eqns. (40) and (42) we obtain a closed equation for $X_o$

$$\frac{1}{X_o} = \left(\frac{1}{1-\lambda X_o}\right)^2 + \rho X_o f_{AB}^2 \xi I_d + \sum_{n=3}^{\infty} \frac{(f_{AB} g_{HS}(\sigma) K)^n}{\sigma^3} (\rho X_o)^{n-1} I_r^{(n)} - (1-\Psi)\left(\frac{\lambda X_o}{1-\lambda X_o}\right)^2 \quad (43)$$

Once Equation (43) is solved for $X_o$, Eq. (40) can be evaluated to obtain $X_A$.

To apply the theory we must evaluate Eqns. (10) and (21) for the integrals $\Psi$ and $\Theta^{(n)}$. Due to the highly discontinuous nature of these integrals they must be evaluated using monte carlo integration. Obviously we cannot evaluate the sum over ring fractions for all possible ring sizes, so we truncate the sum at $n = 7$. As will be seen, this is more than sufficient to describe the conditions studied in this paper. The specific method used to evaluate these integrals can be found in the appendix.

Figure 4 shows numerical calculations for $\Psi$ and $\Gamma^{(n)}$ for $n = 3 - 7$. We have also included the analytical solution of $I_d$ Eq. (12) for comparison. All calculations were performed for the case $\beta_c = 27°$ and $r_c = 1.1\sigma$. As expected $\Psi$ vanishes for small $\alpha_{AB}$ due to steric hindrance and becomes unity for large $\alpha_{AB}$ when association at one patch no longer interferes with the ability of the other patch to bond. The ring integrals $\Gamma^{(n)}$ are peaked around an optimum bond angle for ring formation and the maximums of $\Gamma^{(n)}$ decrease and shift to larger bond angles as $n$ increases. The double bonding integral $I_d$, which represents the probability two colloids are oriented such that double bonding can occur, vanishes for $\alpha_{AB} > 54°$. In the limiting case $\alpha_{AB} = 0°$ the integral $I_d \to \kappa$ due to the fact that since both patches are superimposed the probability two colloids are oriented for double bonding is just the probability that two monomers are oriented such that a specific patch on one colloid can bond to a specific patch on the other. By inspection of the integrals in Fig. 4 we should expect, in strongly associating fluids, double bonded dimers to dominate for small $\alpha_{AB}$, rings to dominate for moderate $\alpha_{AB}$ and chains to dominate for large $\alpha_{AB}$. It will be shown that this is indeed the case.

**III: Simulations**

To test the theory we perform NVT (constant N, V and T) and NPT (constant pressure P, V and T) monte carlo simulations. The colloids interact with the potential given by Eq. (1) with $r_c = 1.1\sigma$ and $\beta_c = 27°$. The simulations were allowed to equilibrate for $10^6$ cycles and averages were taken over another $10^6$ cycles. A cycle is defined as N attempted trial moves where a trial move is defined as an attempted relocation and reorientation of a colloid. For the NPT simulations a volume change was attempted once each cycle.

For the majority of simulations performed in this work, small clusters of associated colloids (double bonded dimers, trimer rings etc…) dominate the fluid even at low temperatures. For this reason a choice of N = 256 colloids is sufficient to obtain good statistics. For larger bond angles where colloids can polymerize into longer chains at low temperatures[29] we performed additional simulations using N = 864 colloids. Increasing the number of colloids had no significant effect on the simulated quantities.

## IV: Results

In this section we compare predictions of the theory to monte carlo simulation results. In part A of this section we compare theory and simulation results when the bond angle is the independent variable. In part B, we hold bond angle and density constant and vary temperature. As will be seen, theory and simulation are in excellent agreement.

### A: Bond angle dependence

We begin with a discussion of the effect of bond angle on the fraction of colloids which are monomers $X_o$, bonded once $X_1$ and bonded twice $X_2 = 1 - X_o - X_1$ at constant packing fraction $\eta$ and association energy $\varepsilon^* = \varepsilon_{AB}/k_BT$. Figure 5 gives these fractions at a relatively low $\varepsilon^* = 5$ and Fig. 6 gives them for the higher association energy case $\varepsilon^* = 8$. For each $\varepsilon^*$ we consider low density $\eta = 0.1$ and high density $\eta = 0.4$ fluids. In all cases $X_2$ dominates for small $\alpha_{AB}$ and decreases to some limiting value as the bond angle dependence saturates around $\alpha_{AB} = 60°$ for $\varepsilon^* = 5$ and $\alpha_{AB} = 90°$ for $\varepsilon^* = 8$. It is at these bond angles that TPT1 becomes accurate. The fractions $X_o$ and $X_1$ are a maximum at large $\alpha_{AB}$ and then decrease rapidly as $X_2$ increases at smaller $\alpha_{AB}$. As expected, the general trend observed for all $\alpha_{AB}$ is that association between the colloids increases with increasing $\varepsilon^*$ and $\eta$. Overall the theory and simulation are in excellent agreement.

To better explain the behavior observed in Figs. 5 – 6 we show the fraction of colloids in doubly bonded dimers $X_d$, fraction in rings of size $n$ $X_n^{ring}$ and fraction of colloids bonded at both patches in a chain (not a ring or double bonded dimer) $X_{2c} = \Psi X_o (X_A \lambda)^2$ for the case $\varepsilon^* = 8$ in Fig. 7. For small $\alpha_{AB}$, double bonded dimers dominate. For these small bond angles $I_d$ is maximum, ring formation is impossible due to vanishing $\Gamma^{(n)}$, $\Psi \to 0$ meaning steric hindrance between patches is nearly complete, and finally there are more mutual orientations where colloids can form a double bond than there are mutual orientations where a single bond is formed (at $\alpha_{AB} = 0°$ single bonding of a patch becomes impossible). The tendency of the colloids to double bond is the genesis of the $X_2$ dominance for small $\alpha_{AB}$. As bond angle is increased the solid angle available for double bonding decreases and vanishes completely at $\alpha_{AB} = 54°$. In the region $50° \leq \alpha_{AB} \leq 70°$ rings become the dominant type of associated cluster in the fluid. The reason for this can be seen in the geometric integrals given in Fig. 4. In this region $\Psi$ is depleted and the ring integrals $\Gamma^{(3)}$ and $\Gamma^{(4)}$ are significant. The maximum of $X_3^{ring}$ is significantly larger than the maximum of $X_4^{ring}$, which in turn is much larger than that seen in $X_5^{ring}$. For $n > 5$ $X_n^{ring}$ ring is small for all $\alpha_{AB}$ at these conditions. For $\alpha_{AB} > 70°$, $X_{2c}$ becomes the dominant contribution to $X_2$ due to decreasing $\Gamma^{(n)}$ and the fact that there is little steric hindrance between patches. The theory does an excellent job in describing the various bonding fractions of the system.

Figure 8 shows the $\alpha_{AB}$ dependence of the excess internal energy $E^* = E/Nk_BT$. For each case $|E^*|$ is largest for small $\alpha_{AB}$. This is due the fact that at these bond angles double bonded dimers dominate which give the energetic benefit of forming a double bond for the entropic penalty of forming a single bond in the large $\alpha_{AB}$ case. Increasing $\alpha_{AB}$ decreases $X_d$ resulting in a corresponding decrease in $|E^*|$. For $\varepsilon^* = 8$, $E^*$ shows oscillatory behavior in the region $40° \leq \alpha_{AB} \leq 80°$ as the system switches between various modes of association. The energy reaches a limiting value near $\alpha_{AB} = 80°$ at which TPT1 would give accurate predictions. As can be seen, the current theory is in excellent agreement with simulation over the full range of bond angles.

Now we wish to explore the effect of $\alpha_{AB}$ on pressure. To determine the performance of the new theory in the prediction of pressure we performed NPT simulations for 3 isotherms. Figure 9 compares theory and simulation predictions for $\alpha_{AB} = 45°$ at $\varepsilon^* = 4$ and 8 and $\alpha_{AB} = 180°$ at $\varepsilon^* = 8$. For $\alpha_{AB} = 45°$ increasing $\varepsilon^*$ decreases pressure due to the fact that more colloids are associating into clusters. The system has a significantly lower pressure for $\alpha_{AB} = 180°$ than $\alpha_{AB} = 45°$ at $\varepsilon^* = 8$. The reason for this can be seen in the types of associated clusters observed in Fig. 7. For $\alpha_{AB} = 45°$ the system is dominated by small clusters such as double bonded dimers and triatomic rings, while for $\alpha_{AB} = 180°$ the system is dominated by larger clusters of associated linear chains. Overall the theory does a good job in predicting the pressure isotherms.

In Fig. 10 we hold density and $\varepsilon^*$ constant and plot reduced pressure over the full bond angle range. For $\eta = 0.1$ and $\varepsilon^* = 5$ the pressure is at a minimum for $\alpha_{AB} = 0°$ where the majority of colloids are associated into double bonded dimers and then increases to a limiting value near $\alpha_{AB} = 50°$ where most colloids are monomers (Fig. 5) at these conditions. Increasing to $\varepsilon^* = 8$ at $\eta = 0.1$ we see the opposite behavior; now pressure is a maximum for $\alpha_{AB} = 0°$ where nearly all colloids are bonded in double bonded dimers, remains relatively constant until $\alpha_{AB} \sim 40°$, goes through a maximum near $\alpha_{AB} \sim 45°$ where $X_2$ goes through a minimum (Fig. 6) and then decreases to a limiting value as the bond angle dependence saturates and the system becomes dominated by linear chains. Increasing the packing fraction to $\eta = 0.4$ we see similar behavior with the exception that for $\varepsilon^* = 5$ the pressure for $\alpha_{AB} = 180°$ is only slightly higher than the $\alpha_{AB} = 0°$ case. This is due to the increase in association at this density resulting in a maximum in pressure near $\alpha_{AB} \sim 45°$ while the maximum for the $\varepsilon^* = 8$ case disappears.

## B: Dependence on $\varepsilon^*$ for fixed bond angle at $\eta = 0.3$

We also study the effect of $\varepsilon^*$ on association for the fixed bond angles of $\alpha_{AB} = 30°$ and $60°$ for $\eta = 0.3$. Figure 11 compares theory and simulation predictions of $X_o$, $X_1$ and $X_2$ for these systems as a function of $\varepsilon^*$. In both cases for small $\varepsilon^*$, $X_o$ dominates due to the fact that the small energetic benefit of forming an association bond does not outweigh the entropic penalty of orienting the two colloids. As $\varepsilon^*$ is increased the fraction $X_1$ increases going through a maximum and then vanishing for large $\varepsilon^*$. The maximum in $X_1$ results from an increase in the $X_2$ which becomes dominant for large $\varepsilon^*$. The maximum in $X_1$ for $\alpha_{AB} = 30°$ occurs at a significantly lower $\varepsilon^*$ than for $\alpha_{AB} = 60°$.

The origin of this behavior can be found in Fig. 12 which shows the significant contributions (dimers, rings, and chains) to $X_2$. For $\alpha_{AB} = 30°$ $X_d$ is the only significant contribution to $X_2$ due to the fact that it is very difficult to form rings at this bond angle. The remaining two possibilities for colloids to become fully bonded is to orient and position multiple colloids to polymerize into chains, or to simply orient and position two colloids to form a double bond. The entropic penalty is much less for the double bonding case resulting in the complete dominance of $X_d$. This dominance of $X_d$ is also the origin of the shift of the maximum of $X_1$ to lower association energies. From Fig. 4 we see the ratio $I_d / \kappa \sim 0.1$ for $\alpha_{AB} = 30°$ meaning the entropic penalty of forming a double bond is only 10 times that of forming a single bond;

however, the energetic benefit of forming the double bond will increase like $f_{AB}$ which becomes very large for high association energies. For this reason $X_1$ is only dominant for a small range for low association energies at the bond angle $\alpha_{AB} = 30°$.

For $\alpha_{AB} = 60°$ double bonding is no longer possible and the significant contributions to $X_2$ are $X_{2c}$, $X_3^{ring}$ and $X_4^{ring}$. This can be seen in the bottom panel of Fig. 12. For low $\varepsilon^*$ the majority of associated colloids are only bonded once (see Fig. 11), meaning the majority of colloids bonded twice are the center colloid in triatomic chains resulting in $X_{2c}$ being the dominant contribution to $X_2$. Increasing $\varepsilon^*$, $X_3^{ring}$ rapidly becomes the dominant type of associated cluster in the fluid forcing a maximum in $X_{2c}$ which becomes very small in strongly associating systems. The fraction $X_4^{ring}$ shows a nearly linear increase with $\varepsilon^*$, overtaking $X_{2c}$ near $\varepsilon^* \sim 9.5$. In both Figs. 11 and 12 theory and simulation are in excellent agreement.

We conclude this subsection with Fig. 13 which compares theory and simulation for the energy $E^*$ at these two bond angles. For both cases $E^* \to -\varepsilon^*$ for large $\varepsilon^*$ and, of course, $E^* \to 0$ for $\varepsilon^* \to 0$; however, at moderate $\varepsilon^*$ we find that $|E^*|$ is larger for $\alpha_{AB} = 30°$ than for $\alpha_{AB} = 60°$. This is to be expected since for moderate association energies $X_2$ is always higher in the $\alpha_{AB} = 30°$ system as compared to the $\alpha_{AB} = 60°$ system. The new theory does an excellent job of predicting the temperature dependence of the internal energy.

## V: Conclusions

We have extended Wertheim's theory to model 2 patch colloids where the patches can be separated by any bond angle $\alpha_{AB}$. We used Wertheim's resummed perturbation theory[8] to account for blockage effects in chain formation, Sear and Jacksons graph for double bonded dimers[19] and modified the ring graph of Sear and Jackson[20] to account for the association of colloids into non overlapping rings. We obtained an analytical solution for the double bonding integral $I_d$ which represents the probability that two colloids are oriented such that double bonding can occur, this quantity was treated as a parameter in previous studies[19]. The integrals $\Psi$ (which account for the fact that bonding at one patch may block bonding at the other) and $\Gamma^{(n)}$ (which are proportional to the number of ways *n* colloids can position themselves to form rings of size *n*) were evaluated using monte carlo integration as a function of $\alpha_{AB}$. This is the first application of Wertheim's theory to associating fluids which explicitly accounts for the effect of bond angle.

The new theory was extensively tested against new monte carlo simulation data and found to be very accurate. It was shown that $\alpha_{AB}$ plays a crucial role in the thermodynamics of these fluids. In systems which exhibit significant association, double bonded dimers dominate for small $\alpha_{AB}$. Increasing $\alpha_{AB}$ further, there is a transition to a ring dominated fluid; increasing $\alpha_{AB}$ even further, ring formation becomes unlikely and the system becomes dominated by associated chains of colloids. In the region $40° \leq \alpha_{AB} \leq 90°$ there is a vicious competition between the various modes of association. The new theory was shown to successfully account for this full range of interactions and accurately predict the fraction of colloids in each type of associated cluster, internal energy and pressure.

The analysis presented in this work is restricted to 2 patch colloids. However, it is known[30] that to have a liquid – vapor phase transition when only 1 bond per patch is allowed a colloid must have a minimum of three patches. In addition, lattice simulations have shown[22] that bond angle can have significant effect on the phase diagram of patchy colloids. To allow for the study of the effect of bond angle on liquid – vapor equilibria the approach developed in this paper must be extended to allow for more than two patches. This will be the subject of a future publication.

In addition to bulk fluids, Wertheim's theory has also found wide application in the study of inhomogeneous associating fluids in the form of classical density functional theory.[18, 31-38] A general extension of the approach presented in this paper to inhomogeneous systems, in the form of classical density functional theory, could provide a valuable tool in the study of inhomogeneous associating fluids.

## **Acknowledgments**

The financial support of The Robert A. Welch Foundation Grant No. C – 1241 is gratefully acknowledged29

**Appendix: Evaluation of $\Theta^{(n)}$**

In this appendix we discuss the numerical evaluation of $\Theta^{(n)}$ given by Eq. (21). The integral $\Theta^{(n)}$ represents the total number of states (positions and orientations for *n* colloids) which lead to rings of size *n*. For a ring to form each colloid must be properly positioned and oriented such that there is a ring of attraction bonds, and there can be no overlap of spheres in the ring. Due to the highly discontinuous nature of this integral we employ a monte carlo integration technique. In monte carlo integration we exploit the mean value theorem which states that any integral *I* can be written as the average value of its integrand multiplied by the volume of integration.[39] For a simple 1 – D integral

$$I = \int_a^b f(x)dx = \bar{f} \times (b-a) \tag{A1}$$

In monte carlo integration the average $\bar{f}$ is evaluated by generating *Q* random numbers *q* and taking the average as

$$\bar{f} = \frac{1}{Q}\sum_{q=1}^{Q} f(q) \tag{A2}$$

In the limit $Q \to \infty$ the integral given by Eq. (A1) becomes exact.

We evaluate $\Theta^{(n)}$ using this method. Figure 14 shows a cluster of 4 colloids associated in a 4 – mer ring. We have exaggerated the distances between colloids for clarity. The orientation of each colloid *j* is defined by three angles $\{0 \leq \phi_j \leq 2\pi\}$, $\{0 \leq \gamma_j \leq 2\pi\}$ and $\{-1 \leq \cos\theta_j \leq 1\}$ in the convention used in Goldstein.[40] From these three orientation angles the site orientation vectors $\vec{r}_A^{(j)}$ and $\vec{r}_B^{(j)}$ are defined for colloid *j*.

To generate a chain configuration we fix the first colloid in the chain at the origin of a global coordinate system and give the colloid some convenient orientation $\Omega_1$. We generate colloid 2 in a spherical coordinate system centered on colloid 1 whose $z$ axis lays parallel to the vector $\vec{r}_B^{(1)}$ and generate an orientation for colloid 2 in an orientation reference frame whose $z$ axis is parallel to $-\vec{r}_B^{(1)}$. With these choices of reference frames it is possible to generate a position $\vec{r}'_{12}$ and orientation $\Omega'_2$ such that it is guaranteed the bonding constraints of the potential Eq. (1) are satisfied. To obtain the actual position $\vec{r}_{12} = \vec{r}_2 - \vec{r}_1$ and orientation $\Omega_2$ we simply rotate back into the global position and orientation reference frames. To generate the third colloid we follow the same procedure with a coordinate system centered on colloid 2 whose z axis is parallel to $\vec{r}_B^{(2)}$ for the position reference frame and $-\vec{r}_B^{(2)}$ for the orientation reference frame. We continue this process until we have generated a position and orientation for each colloid. This is our number $\vec{q}$ as discussed previously, although now $\vec{q}$ is a vector which describes a chain configuration of $n$ associated colloids.

Following this approach of generating configurations $\vec{q}$, the integral $\Theta^{(n)}$ is written as

$$\Theta^{(n)} = \left(8\pi^2 K\right)^{n-1} \times \overline{f(\vec{q})} \tag{A3}$$

Where the function $f(\vec{q})$ is given by

$$f(\vec{q}) = \frac{\phi_{AB}(1,n)}{-\varepsilon_{AB}} \left( \prod_{\substack{\text{all pairs} \\ \{l,k\}}} e_{HS}(r_{lk}) \right) \prod_{m=2}^{n} \left( \frac{r_{m,m-1}}{\sigma} \right)^2 \tag{A4}$$

To evaluate Eq. (A3) would use $\sim 10^8$ chain conformations $\vec{q}$ at each bond angle and ring size.

We find that the integrals $\Gamma^{(n)} = \sigma^3 \overline{f(\vec{q})}/K$ (given by Eq. (23)) are well correlated as a function of bond angle using the skewed Gaussian function

$$\Gamma^{(n)} = A_n \exp\left(-B_n(\alpha_{AB} - C_n)^2\right)\left(1 + erf\left(-D_n(\alpha_{AB} - C_n)\right)\right). \tag{A5}$$

The constants $A_n, B_n, C_n$ and $D_n$ depend on ring size $n$ and are given for ring sizes $n = 3 - 10$ in table 1. In Eq. (A5) $\alpha_{AB}$ must be given in degrees.

**Table Captions:**

**Table 1:** Constants for ring integral correlation Eq. (A5) for ring sizes $n = 3 – 10$

**Figure captions:**

**Figure 1:** Diagram of two patch colloid with patches separated by the bond angle $\alpha_{AB}$. The angle $\beta_c$ defines the size of the patches

**Figure 2:** Types of associated clusters accounted for in the theory. Here we depict 4-mer rings, but the theory accounts for all ring sizes

**Figure 3:** Diagram of double bonded colloids. The line connecting the centers of each colloid must pass through the solid angle of the overlap of both patches given by Eq. (13), outlined in dashed curve, for double bonding to occur

**Figure 4:** Geometric integrals used in the application of the theory. $I_d$ is the probability two colloids are oriented such that double bonding can occur and is given by Eq. (12), $\Psi$ is the blockage parameter given by Eq. (10) which accounts for the fact bonding at one patch can interfere with bonding at the other, $\Gamma^{(n)}$ are the ring integrals given by Eq. (23) which are proportional to the total number of ways $n$ colloids can associate into rings of size $n$. Inset of figure gives $\Gamma^{(n)}$ for $n = 5 – 7$ with 5 being the largest peak and 7 being the smallest

**Figure 5:** Monomer fractions $X_o$ (short dashed line – theory, triangles – simulation), fractions of colloids bonded once $X_1$ (long dashed line – theory, squares – simulation) and fractions of colloids bonded twice $X_2$ (solid line – theory, circles – simulation) as a function of bond angle $\alpha_{AB}$ for $\varepsilon^* = 5$ at packing fractions of $\eta = 0.1$ (top) and $\eta = 0.4$ (bottom)

**Figure 6:** Same as Fig. 5 with $\varepsilon^* = 8$

**Figure 7:** Fractions of colloids bonded twice in dimers $X_d$, twice in chains $X_{2c}$ and twice in rings of size $n$ $X_n^{ring}$ for $n = 3 - 5$ for an association energy of $\varepsilon^* = 8$ at packing fractions $\eta = 0.1$ (top) and $\eta = 0.4$ (bottom). Curves give theoretical predictions and symbols give simulation results. Insets show $X_4^{ring}$ (triangles) and $X_5^{ring}$ (crosses)

**Figure 8:** Bond angle dependence of the excess free energy per colloid $E^* = E/Nk_BT$ for $\varepsilon^* = 5$ (solid curve – theory, circles – simulation) and $\varepsilon^* = 8$ (dashed curve – theory, triangles – simulation) for $\eta = 0.1$ (top) and $\eta = 0.4$ (bottom)

**Figure 9:** Pressure isotherms for $\alpha_{AB} = 45°$ at $\varepsilon^* = 4$ (solid line – theory, filled circles – simulation) and $\varepsilon^* = 8$ (short dashed line – theory, open circles – simulation). Long dashed line and open triangles give theory and simulation predictions for the case $\alpha_{AB} = 180°$ and $\varepsilon^* = 8$

**Figure 10:** Bond angle dependence of pressure $P^* = P\sigma^3/k_BT$ at association energies $\varepsilon^* = 5$ (solid curve) and $\varepsilon^* = 8$ (dashed curve) for packing fractions $\eta = 0.1$ (top) and $\eta = 0.4$ (bottom)

**Figure 11:** $\varepsilon^*$ dependence (inverse temperature) of fractions $X_o$ (short dashed curve – theory, triangles – simulation), $X_1$ (long dashed curve – theory, squares – simulation) and $X_2$ (solid curve – theory, circles – simulation) at a packing fraction $\eta = 0.3$ for $\alpha_{AB} = 30°$ (top) and $\alpha_{AB} = 60°$ (bottom)

**Figure 12:** Significant components of the fractions $X_2$ presented in Fig. 11. Top panel gives the only significant contribution $X_d$ (solid curve – theory, diamonds – simulation) for $\alpha_{AB} = 30°$.

Bottom panel gives relevant contributions for $\alpha_{AB} = 60°$: $X_3^{ring}$ (solid curve – theory, squares – simulation), $X_4^{ring}$ (long dashed curve – theory, crosses – simulation) and $X_{2c}$ (short dashed curve – theory, triangles – simulation)

**Figure 13:** $\varepsilon^*$ dependence of the excess free energy per colloid $E^* = E/Nk_BT$ for $\alpha_{AB} = 30°$ (dashed curve – theory, squares – simulation) and $\alpha_{AB} = 60°$ (solid curve – theory, triangles – simulation) at a packing fraction $\eta = 0.3$

**Figure 14:** Four colloids associated into a 4 – mer ring. Distances between colloids are exaggerated for clarity.

**Table 1:**

| $n$ | $A_n$ | $B_n$ | $C_n$ | $D_n$ |
|---|---|---|---|---|
| 3 | 0.681 | 0.00514 | 60.0 | 0.00116 |
| 4 | 0.0651 | 0.00111 | 94.4 | 0.0907 |
| 5 | 0.0231 | 0.00159 | 105.8 | 0.0668 |
| 6 | 0.0111 | 0.00180 | 112.0 | 0.0461 |
| 7 | 0.00363 | 0.00278 | 107.5 | 0.0452 |
| 8 | 0.00248 | 0.00271 | 109.8 | 0.0452 |
| 9 | 0.00208 | 0.00291 | 112.2 | 0.0452 |
| 10 | 0.00165 | 0.00276 | 113.6 | 0.0452 |

**Figure 1:**

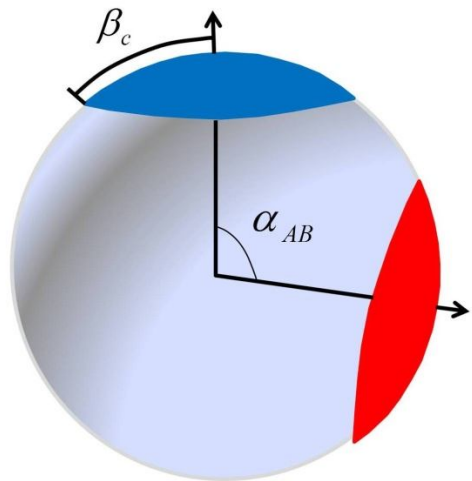

**Figure 2:**

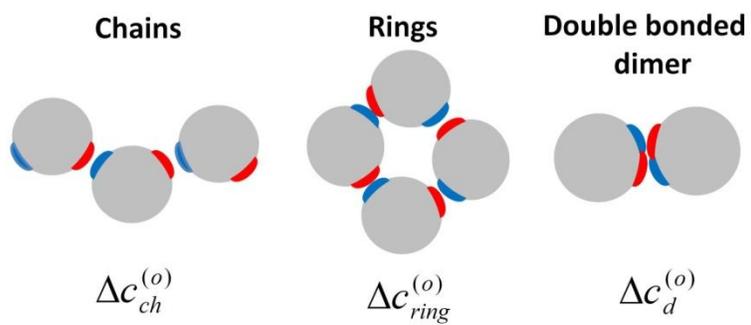

**Figure 3:**

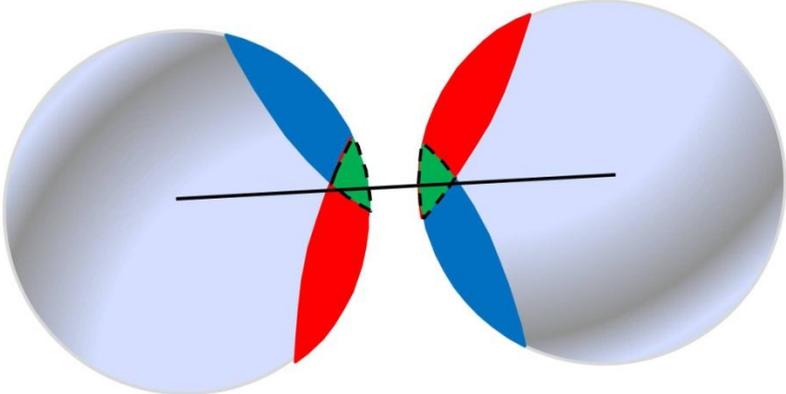

**Figure 4:**

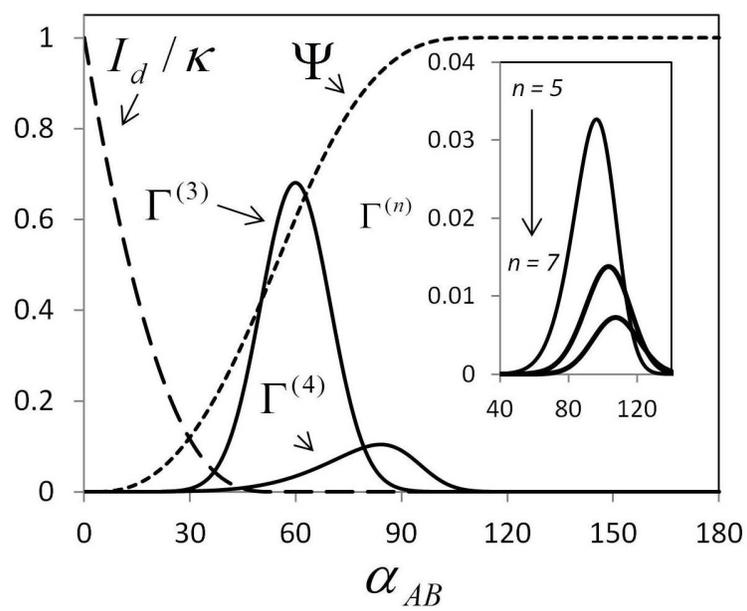

**Figure 5:**

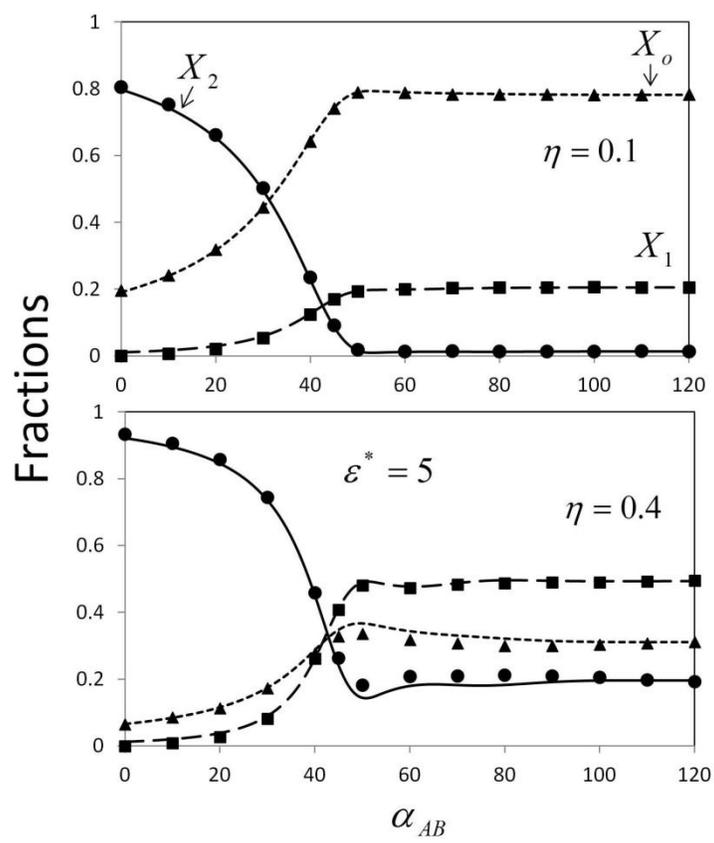

**Figure 6:**

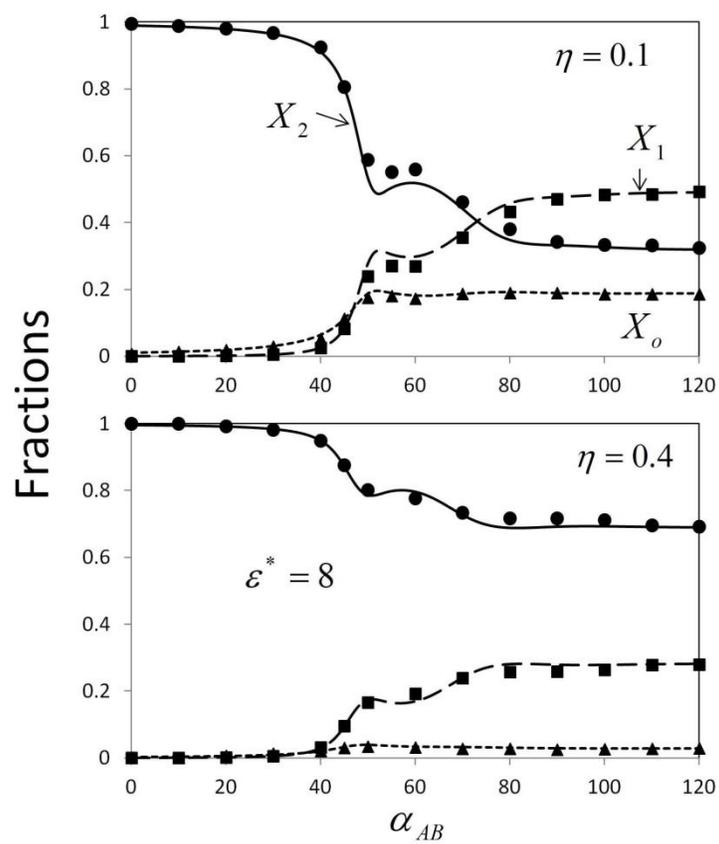

**Figure 7:**

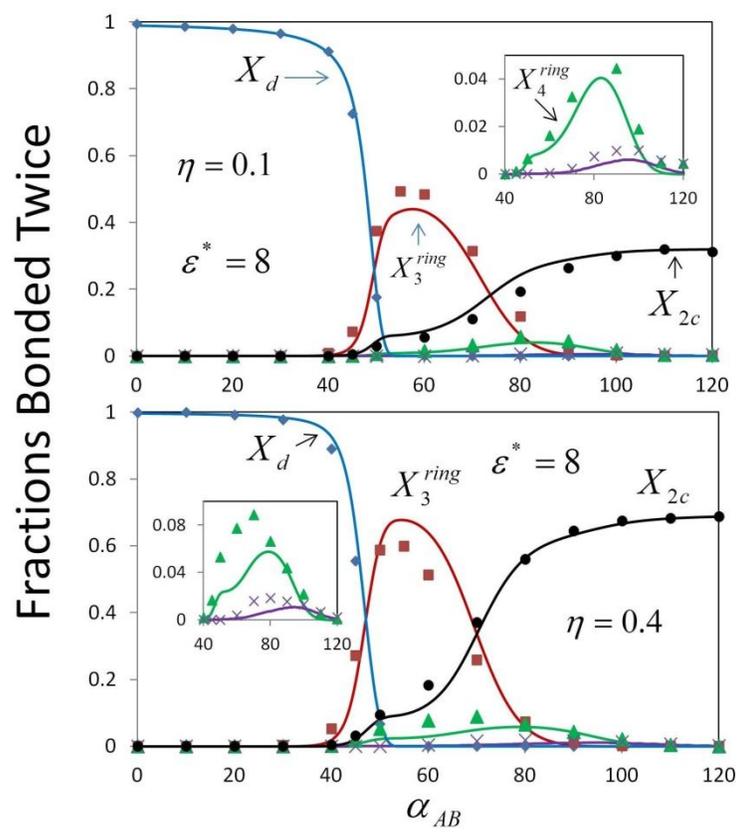

**Figure 8:**

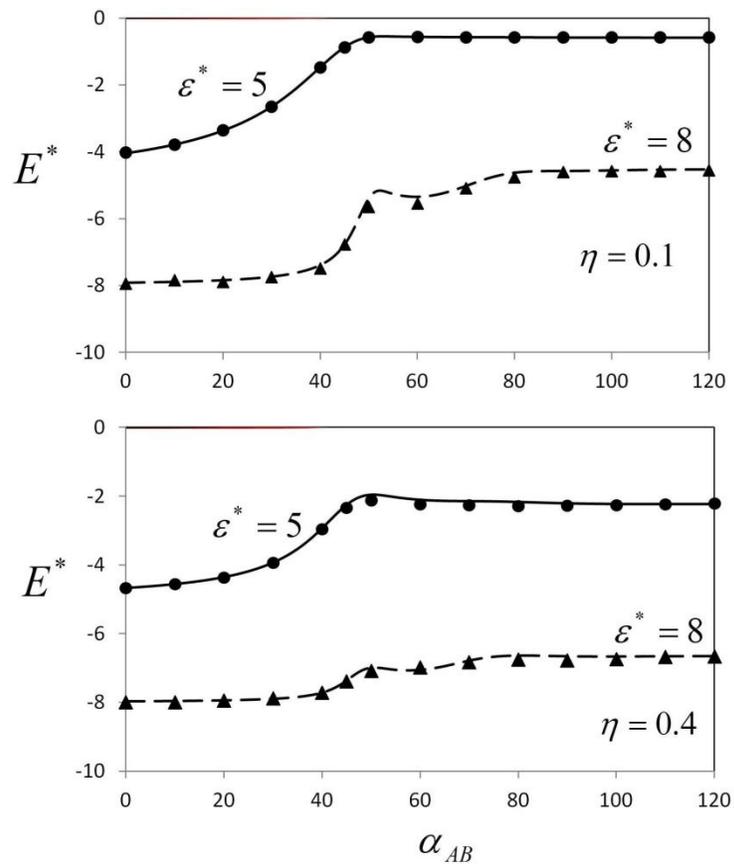

**Figure 9:**

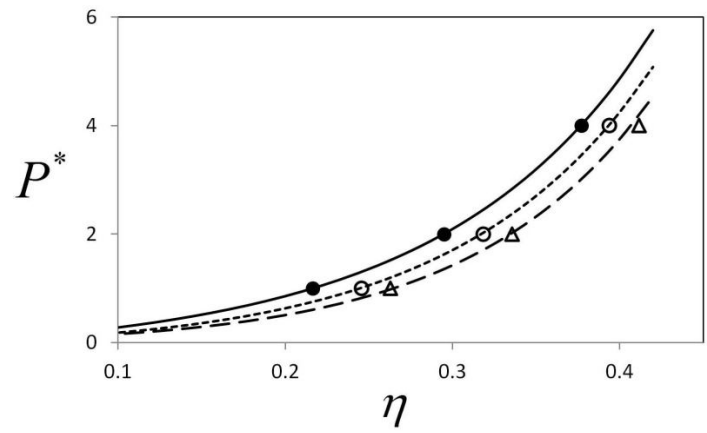

**Figure 10:**

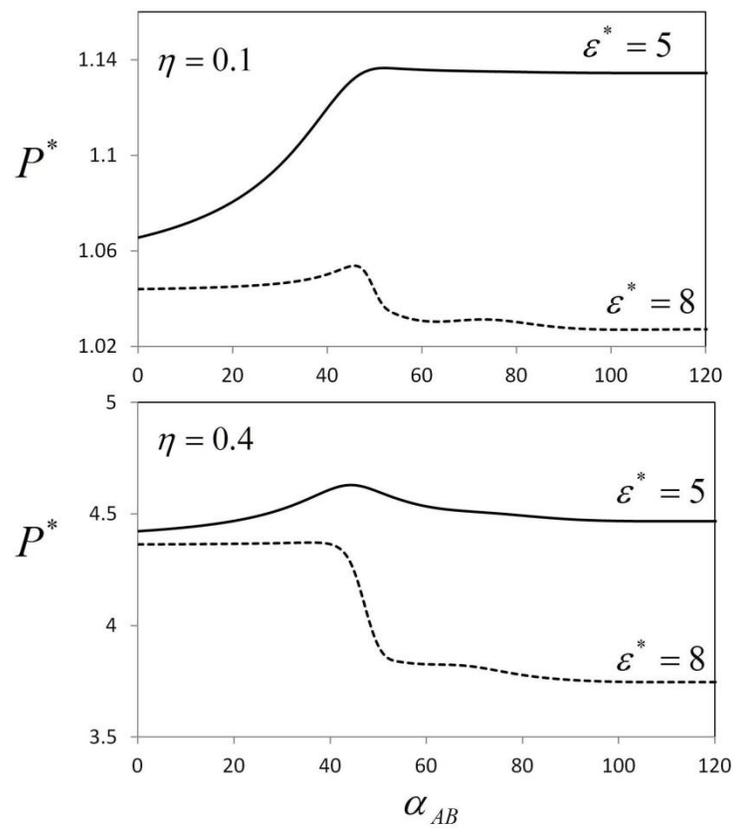

**Figure 11:**

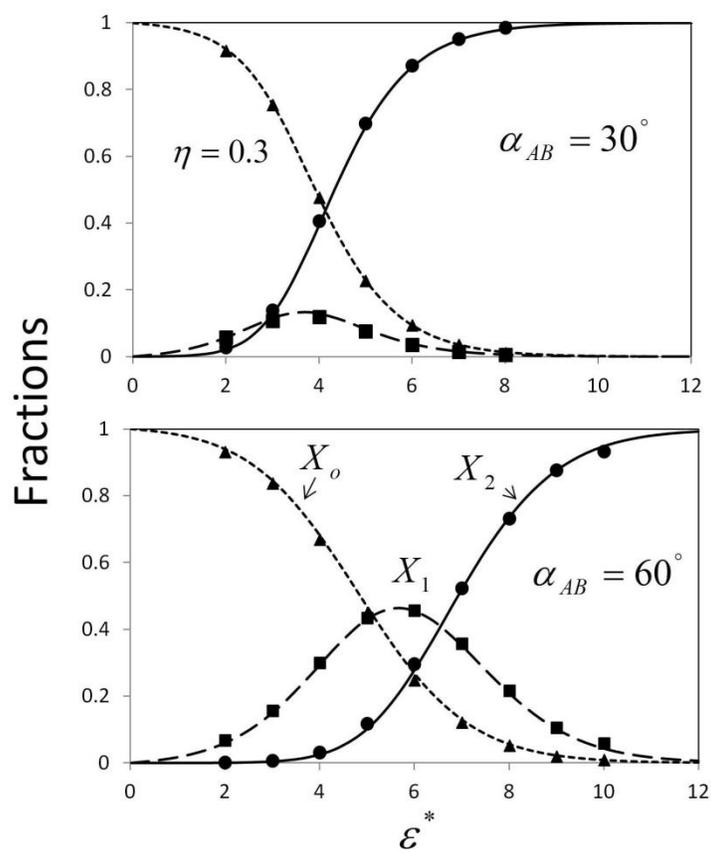

**Figure 12:**

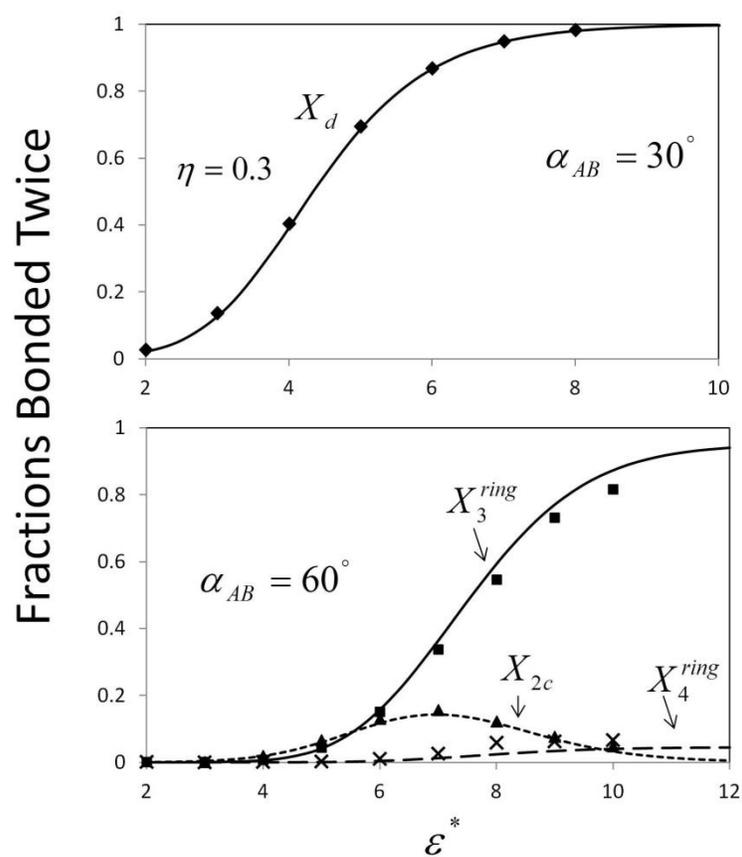

**Figure 13:**

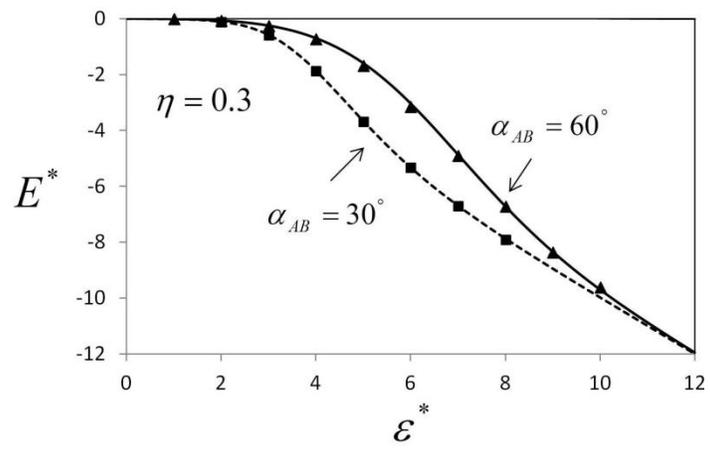

**Figure 14:**

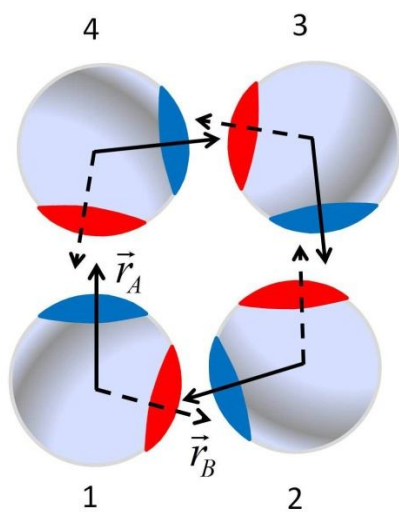